\documentclass[10pt,twocolumn,oneside,submit]{JCNtran}

\usepackage[dvips]{epsfig}
\usepackage[latin1]{inputenc}
\usepackage[T1]{fontenc}
\usepackage{float}

\def\BibTeX{{\rm B\kern-.05em{\sc i\kern-.025em b}\kern-.08em
    T\kern-.1667em\lower.7ex\hbox{E}\kern-.125emX}}

\begin{document}
\bibliographystyle{jcn}

\title{An Adaptive MAC Protocol for Wireless LANs}
\author{Amin Jamali, Seyed Mostafa Safavi Hemami, Mehdi Berenjkoub,
and Hossein Saidi
\thanks{Manuscript received 23 September 2012; revised 2 November 2013.}
\thanks{A.~Jamali, M.~Berenjkoub, and H.~Saidi are with the Electrical and Computer Engineering Department, Isfahan University of Technology, Isfahan, Iran, emails: a.jamali@ec.iut.ac.ir;brnjkb@cc.iut.ac.ir;hsaidi@cc.iut.ac.ir. 
}\thanks{S. M.~Safavi Hemami is with the Department of Electrical Engineering, Amirkabir University of Technology, Tehran, Iran, e-mail: msafavi@aut.ac.ir.}} \markboth{JOURNAL OF
COMMUNICATIONS AND NETWORKS, VOL. 5, NO. 3, SEPTEMBER
2014}{Jamali \lowercase{\textit{et al}}.: An Adaptive MAC Protocol for Wireless LANs} \maketitle

\begin{abstract}
This paper focuses on contention-based Medium Access Control (MAC) protocols used in Wireless Local Area Networks (WLANs). We propose a novel MAC protocol called Adaptive Backoff Tuning MAC (ABTMAC) based on IEEE 802.11 DCF. In our proposed MAC protocol, we utilize a fixed transmission attempt rate and each node dynamically adjusts its backoff window size considering the current network status. We determined the appropriate transmission attempt rate for both cases where the Request-To-Send/Clear-To-Send (RTS/CTS) mechanism was and was not employed. Robustness against performance degradation caused by the difference between desired and actual values of the attempt rate parameter is considered when setting it. The performance of the protocol is evaluated analytically and through simulations. These results indicate that a wireless network utilizing ABTMAC performs better than one using IEEE 802.11 DCF.
\end{abstract}


\section{\uppercase{Introduction}}
\label{sec:introd}
\par For decades, Ethernet has been the main network technology for local area networks (LANs). Recently, there has been a rapid development in the field of wireless communications and as a result, WLANs have emerged as a dominant means of wireless communications and Internet access. Due to their low cost, ease of deployment, and mobility support, IEEE 802.11 WLANs have been widely used and are now the dominant WLAN technology.
\par Two types of coordination are proposed in the IEEE 802.11 standard: Distributed Coordination Function (DCF) and Point Coordination Function (PCF). DCF is widely utilized in current WLANs and is intended for distributed, contention-based, asynchronous access to a channel, while PCF has been proposed for contention-free and centralized access. PCF is intended to support real-time services (by using a centralized polling mechanism), but it is not generally supported by the current Network Interface Cards (NICs).
\par A well-designed MAC protocol should provide some specific features. The performance metrics of interest include throughput, fairness, and packet transmission delay as well as priority in an environment that supports multiple services. Several MAC layer protocol capacity enhancement techniques for the IEEE 802.11 standard exist in the literature [1]-[5]. These techniques include introducing delay before packet transmission, dynamically adjusting backoff window size, slot reservations, and cross-layer design of protocols. However, most of the previous methods do not consider stability issues of the network. In our proposed MAC protocol ABTMAC that is based on IEEE 802.11 DCF, each node dynamically adjusts its backoff window size according to the current network status. Protocol parameters are selected in ABTMAC to consider WLAN stability.
\par An analytical model that approximates the time until the next transmission attempt in IEEE 802.11 DCF as a random variable with an exponential distribution was presented in [1]. In this model, a station with a pending frame transmits this frame with a specific transmission attempt rate at a given time. It should be noted that the attempt rate of stations is variable in legacy DCF, and the attempt rate becomes higher as the number of contending stations increases. To the best of our knowledge, none of the previous studies in this area of research have considered the problem of finding an appropriate transmission attempt rate for high performance transmission in a wireless network. In this paper, we determine an appropriate transmission attempt rate for a WLAN. Stations set their transmission attempt rates to this value instead of selecting contention window (CW) sizes that increase the attempt rate as the network population grows. Each station calculates the optimal backoff time for the transmission of its pending frame by applying the fixed transmission attempt rate and the station's estimation of the number of active nodes. ABTMAC uses the carrier sense multiple access with collision avoidance (CSMA/CA) mechanism and performs exponential backoff with the same maximum value for CW size as IEEE 802.11. However, minimum CW size is determined adaptively in ABTMAC. The minimum changes required for the implementation of ABTMAC lead to the backward compatibility of ABTMAC with IEEE 802.11 DCF.
\par Similar to IEEE 802.11 DCF, the exchange of RTS/CTS packets between the transmitter and the receiver before the actual transmission of data packets is allowed in ABTMAC. The attempt rate is determined for both using and not using the RTS/CTS mechanism. When the RTS/CTS mechanism is utilized, the attempt rate is determined considering the stability of the access time and robustness of the protocol to the performance degradation caused by the difference between the actual and desired attempt rates. This difference results from the active nodes' estimation error and nondeterministic backoff times. When the RTS/CTS mechanism is not used, we have developed a method for jointly selecting the attempt rate and packet lengths. Similar to when RTS/CTS is used, the stability and robustness of the protocol are considered when choosing the attempt rate. Furthermore, a predetermined level of Quality of Service (QoS) can be guaranteed for users by using the proposed MAC protocol and choosing different attempt rates for these users. The effectiveness of ABTMAC is investigated analytically and through simulations. The performance evaluation shows that ABTMAC successfully provides a high capacity MAC layer service for higher layers of the network protocol stack. 
\par The remainder of this paper is organized as follows. Section II briefly describes 802.11 DCF and some related previous work. The ABTMAC protocol is detailed in Section III. Section IV includes analytical and simulation results. Section V concludes the paper.

\vspace{10pt}
\section{\uppercase{Preliminaries}}
\label{sec:prilim}
\subsection{IEEE 802.11}
\par The IEEE 802.11 standard [6] includes both the physical (PHY) and MAC layers of wireless networks. A network can be configured in infrastructure-based or ad hoc modes. In an infrastructure-based mode, the network is ``structured'' and hosts communicate by an Access Point (AP). In an ad hoc network, nodes establish a dynamic network and there is no exact structure. The PHY layer specifications may vary in different drafts of the standard. The IEEE 802.11 MAC protocol provides two access methods, DCF, which is mandatory, and PCF, an optional mechanism. In addition, the standard includes the RTS/CTS mechanism to resolve the hidden station problem.

\par As DCF is the basic access method in both wireless infrastructure-based and infrastructure-less environments, we describe the IEEE 802.11 protocol that implements it. DCF is based on the CSMA/CA protocol. In this access method, each node with a packet to transmit senses the channel to find out whether it is in use. If the channel is sensed to be idle for an interval greater than the distributed interframe space (DIFS), the node starts its transmission. If the channel is sensed to be busy, the node defers transmission until the end of the ongoing transmission. The node then initializes its backoff timer with a randomly selected backoff interval and decrements its timer every slot time it senses the channel to be idle. When the channel becomes busy, the station freezes its timer and restarts decrementing after the channel becomes idle for a DIFS again. The node transmits its frame whenever the timer value reaches zero. The receiver transmits the acknowledgment (ACK) after successfully receiving the data frame and waiting for an interval called the short interframe space (SIFS) interval. If the transmitting station does not receive the ACK frame, it will assume that a collision has occurred and will update its CW. After a collision, the CWs of the colliding stations will be multiplied by two and when the CW size reaches a predetermined constant (1024), it will not be increased further. When the number of retransmissions for a frame exceeds a predefined constant (seven as a default), that frame will be dropped.

\subsection{Previous Work}
Various analytical models for IEEE 802.11 DCF can be seen in [1],[2],[5],[7]-[9]. Kim and Hou [1] developed a model-based frame scheduling scheme (MFS) after deriving an analytical model. In MFS, each node keeps track of the number of collisions and time interval between its two consecutive successful transmissions. It then determines the number of currently active nodes, calculates the network utilization with the throughput model, and computes a scheduling delay during which it will not access the wireless medium. MFS is placed on top of IEEE 802.11, and therefore, a new layer is inserted in the network protocol stack. One of the advantages of MFS is that there is no change in the standard IEEE 802.11 DCF. Cal{\`i} et al. [2] analytically derived the average size of the CW that maximizes throughput. Bononi et al. [4] propose a mechanism called Asymptotic Optimal Backoff (AOB) that optimizes IEEE 802.11 during runtime via measuring the network contention level and dynamically adapting backoff window size. However, the RTS/CTS mechanism of DCF was not considered in [2] and [4]. The objectives of our work are similar to the objectives in these studies, nevertheless, the authors did not consider system stability issues in their designs. In addition to improving performance metrics such as throughput and delay, we focus on stability when choosing system parameters.

Hoefel [5] proposed an analytical MAC and PHY cross-layer model to estimate the saturation throughput of IEEE 802.11 WLANs with MAC improvements. Modification of the IEEE 802.11 DCF MAC protocol was accomplished in [5] to support concatenation and multiframe transmission techniques. However, utilizing cross-layer information is one of the disadvantages of this work. Gentle DCF (GDCF) is another mechanism proposed for decreasing collision probability [10]. In this mechanism, the CW is divided by two after $c$ consecutive successful transmissions (instead of resetting the contention level after each successful transmission as in legacy DCF). GDCF may cause unfairness in access to medium for some values of $c$.

In [3], the authors proposed a distributed reservation-based MAC protocol called Early Backoff Announcement (EBA). In this protocol, each station announces its next backoff interval to other stations via the MAC header of the frame it is transmitting. All the stations receiving this information avoid collisions by excluding the same backoff duration when selecting their future backoff value. However, the information about backoff times may be used by Denial of Service (DoS) attackers at the MAC layer to launch a more intelligent and efficient jamming attack. Krishnan and Zakhor [11] showed that an estimate of the probability of collision can be used to increase throughput via link adaptation in 802.11 networks with hidden terminals. This work uses a cross-layer approach. In [12], the authors proposed the Enhanced Grouping-based Distributed Coordinated Function (E-GDCF) scheme to reduce fixed overheads in 802.11 DCF. The advantage of E-GDCF is that it reduces the minimum CW of a WLAN lowering delay in the network.

Concurrent Transmission MAC (CTMAC) [13] is a MAC protocol that supports concurrent transmission. CTMAC inserts an additional control gap between the transmission of control packets (RTS/CTS) and data packets (DATA/ACK), allowing a series of RTS/CTS exchanges to take place between the nodes in the vicinity of the transmitting or receiving node to schedule possible multiple, concurrent data transmissions. In addition, to isolate the possible interference between the DATA and ACK packets, a new ACK sequence mechanism was proposed by the authors. CTMAC works with single-channel, single-transceiver and single-transmission power, and hence its implementation is simple. CONTI [14], which attempts to resolve contention in CONstant TIme, is a MAC scheme that tries to resolve contention in the same number of slots every time. CONTI is a good choice for systems that are intended to offer low-jitter services. In [15], the authors proposed a distributed algorithm that adaptively adjusts the CW configuration of the WLAN. Their work is based on multivariable control theory. Each station uses locally available information to drive the collision probability in the WLAN to an optimal value. This paper also does not consider the RTS/CTS mechanism that is widely deployed in WLANs. Stability of the system is considered in [15]. Stability of throughput and delay in cooperative access [16], maximizing differentiated throughput [17], and cooperative MAC protocol using active relays [18] are the subjects of some other related studies in the field of wireless networks.

\vspace{10pt}
\section{\uppercase{PROPOSED ABTMAC PROTOCOL}}
\label{sec:prop}
Based on the model for IEEE 802.11 DCF introduced in [1], stations transmit their frames with a specific transmission attempt rate in an IEEE 802.11-based network. This attempt rate is different for various network configurations. In fact, it has a larger value for a network with more active stations. A higher attempt rate in a network leads to a higher collision probability. In ABTMAC, stations fix their attempt rate to a specific value and this rate is not increased when the number of active nodes is increased. The mechanism for fixing the attempt rate is described in this section. In addition, appropriate values for the attempt rate are determined for the cases with and without the RTS/CTS mechanism. The attempt rate parameter of an IEEE 802.11 wireless network is introduced in the next paragraph.

In the $p$-persistent model of IEEE 802.11 DCF [2], stations transmit their pending frame in each slot time with a probability of $P_t$. When $P_t$ is equal to $1/(\bar{b}+1)$, the $p$-persistent model closely approximates IEEE 802.11 DCF with an average backoff time of $\bar b$ [2]. In this model, we have $E[CW]=(2/P_t)-1$, where $E[CW]$ is the average CW size [2]. Moreover, $E[CW]<CW_{min}.2^{KlogM}$ is the relation between the average size of the CW, the number of active nodes $M$, and the initial size of the CW denoted by $CW_{min}$ [19]. Constant $K$ is an arbitrary constant ($K>0$), and is set to 1 in our work. The transmission probability of a backlogged node is determined by the backoff timer. The probability that there is no transmission activities in the network is $(1-P_t)^M$. Assuming a large enough value for $M$ and using the approximation $(1-x)^y\approx e^{-xy}$, the time until the next transmission attempt can be approximated as a random variable with an exponential distribution. The rate of this random variable depends on the current set of backoff windows. The attempt rate at a given time is [1]\\
\\
$\lambda (t)=\sum_{i=1}^M {1\over B_i(t)}$,\\
\\
where $B_i(t)$ is the current backoff value of node $i$. Computing the average backoff time as $\bar b=E[B_i(t)]$, the average attempt rate $\lambda $ is given by [1]\\
\\
$\lambda = {{M} \over {\bar b}}$. \hfill(1) \\
\par In legacy DCF, the attempt rate is dependent on the number of active nodes in the network and is between 0.56-1.71 (1/slots) for $M$ between 10-100. The attempt rate is increased automatically when the number of contending nodes grows. An analytical model based on the above assumptions was introduced in [1]. Here, we review some of equations found in this reference. Two analytical components were defined in [1]: the {\it fluid chunk} and {\it MAC fluid}. A {\it fluid chunk} (the time it takes to successfully transmit a frame) is the frame service time and consists of zero or more collision periods followed by a successful frame transmission. A {\it MAC fluid} is composed of a sequence of consecutive fluid chunks. The random variable $N_c$ represents the number of collisions between two consecutive successful transmissions. The $z$ transform of $N_c$ is $N_c(z)$ and we have [1]\\
\\
$N_c(z)=\sum_{n=0}^\infty P[N_c=n].z^n={\lambda e^{-\lambda }\over (1-e^{-\lambda })-(1-e^{-\lambda }-\lambda e^{-\lambda })z}$. \hfill(2) \\
\\
The average number of collisions between two consecutive successful transmissions $\bar n$ is derived as [1]\\
\\
$\bar n=N_c^{(1)}(1)={1-e^{-\lambda }-\lambda e^{-\lambda }\over\lambda e^{-\lambda }}$. \hfill(3) \\
\par Let $\bar x$ be the average time it takes to successfully transmit a frame, $\bar c$ the average length of the collision period, and $\bar {cw}$ the average number of idle slots before a collision or successful transmission. For the case with the RTS/CTS mechanism we have [1]\\
\\
$\bar c =\bar {cw} +tRTS+EIFS$ and\\
$\bar x =\bar {cw} +tRTS+tCTS+tACK+\bar {x'} +DIFS+3SIFS$.\\
\\
Additionally,\\
\\
$\bar c =\bar {cw} +\bar {x'} +EIFS$ and\\
$\bar x =\bar {cw}+tACK+\bar {x'}+DIFS+SIFS$\\
\\
are valid when the RTS/CTS mechanism is not used [1], where $\bar x'$ is the average packet length. We assume that the random variables $L$ and $I$ denote the length of a MAC fluid and the length of an idle period between two consecutive MAC fluids, respectively. The averages of $L$ and $I$ are $\bar l$ and $\bar i$, respectively, and are expressed as [1]\\
\\
$\bar l = {\bar x+\bar n.\bar c\over1-\lambda \bar {cw}(\bar n+1)}$ \hfill(4) \\
\\
and\\
\\
$\bar i = 1/\lambda + DIFS$. \hfill(5) \\
\\
The average number of frame service times in a MAC fluid is $\bar l/(\bar x+\bar f)$, where $\bar f=\bar n.\bar c$ is the total collision period in a frame service time. Considering this fact, the expected throughput is given by [1]\\
\\
$T = {\bar x'.\bar l/(\bar x+\bar f)\over\bar l+\bar i}$. \hfill(6) \\
\par ABTMAC is based on IEEE 802.11 DCF and performs exponential backoff with the same maximum value for CW size, as in the standard IEEE 802.11. However, minimum CW size is determined adaptively in ABTMAC. This MAC protocol is described as follows:\\
\begin{itemize}
\item
Each station estimates the number of active nodes in the network ($\tilde M$) or acquires it from the AP;
\item
The station calculates the average backoff time $\bar b$ by substituting an appropriate transmission attempt rate ($\lambda $) and $\tilde M$ in (1);
\item
The value of $P_t$ is obtained from $P_t=1/(\bar b+1)$;
\item
The average CW size is calculated from $E[CW]=(2/P_t)-1$;
\item
The minimum CW size is set to $E[CW]/2^{logM}$;
\item
When the value of calculated initial CW size exceeds $CW_{max}$ (i.e., 1024), stations use 1024 as $CW_{min}$. Other system parameters and procedures are the same as the IEEE 802.11 standard;
\item
Exponential backoff is performed the same as IEEE 802.11 DCF using $CW_{min}$ and $CW_{max}$ determined above until the successful transmission of the pending frame.\\
\end{itemize}
For example, when the attempt rate is 0.5 (1/slots) and $M$ is 10, $CW_{min}$ will be 20. The value of $CW_{min}$ is increased to 50 for the same attempt rate and 100 active stations. Using the above $CW_{min}$ leads to the transmission of packets with the calculated transmission probability. It should be noted that it is necessary for all of the stations in a WLAN to use the same $K$. Simulation results in Section IV show that using $K=1$ in $E[CW]<CW_{min}.2^{KlogM}$ is appropriate. Network designers are allowed to use other values in their designs. 

An AP can announce the number of active nodes in an infrastructure-based network. In addition, the estimation of the number of active nodes can be done by measuring $E[N_c ]=\bar n$ and substituting it into $\tilde M=10^{E[N_c]/K'}$ where $K'$ is an arbitrary constant [1]. It is evident that finding an appropriate attempt rate is critical. In the following two subsections, we determine this value for achieving high capacity in two modes, both with and without the RTS/CTS mechanism. Furthermore, since the transmission attempt rate determines the average backoff window size, we can guarantee different levels of QoS by choosing different transmission attempt rates for different users or different applications. A more detailed description of QoS differentiation is presented in the following two subsections. Additionally, our proposed MAC protocol is backward compatible with IEEE 802.11 DCF, and 802.11 stations can coexist with nodes that utilize ABTMAC. However, it is evident that stations transmitting their packets using legacy DCF can cause variations in the aggregate attempt rate of the network.

The IEEE 802.11 system parameters used in this paper are given in Table 1. It should be noted that although we have used a specific set of parameters in Subsections III.A and III.B, the method of determining the attempt rate is the same for other sets of parameters and the general form of graphs is preserved.

\begin{table}
\caption{IEEE 802.11 system parameters}
\label{tab:tab1}
\begin{center}
{\small \begin{tabular}{|c|c|}\hline
Channel Rate & 1 Mb/s\\ \hline
Slot Time & 20 $\mu $s\\ \hline
SIFS & 10 $\mu $s\\ \hline
DIFS & 50 $\mu $s\\ \hline
EIFS & SIFS+Phy preamble \& header+tACK+DIFS\\ \hline
Phy preamble & 144 bits\\ \hline
Phy header & 48 bits\\ \hline
MAC header & 224 bits\\ \hline
ACK & 112 bits\\ \hline
RTS & 160 bits\\ \hline
CTS & 112 bits\\ \hline
\end{tabular}}
\end{center}
\end{table}

\subsection{Determination of the attempt rate for the case with the RTS/CTS mechanism}
As we expressed earlier in this paper, nodes are required to fix their average transmission attempt rate in ABTMAC. The attempt rate influences the performance of the system. Therefore, the attempt rate should be chosen carefully. In this subsection, an appropriate value for the attempt rate is determined for the case with the RTS/CTS mechanism.

\par Using (4), (5), and (6) we have\\
\\
$T = {\bar x'\over\bar x+\bar f+1/\lambda +DIFS(1-\lambda \bar {cw}(\bar n+1))-\bar {cw}(\bar n+1)}$. \hfill(7) \\
\\
Substituting the numerical values of the system parameters into (7), we can write it as follows:\\
\\
$T = {\bar x'\over\bar x'+1/\lambda +23.7\bar n+23.2}$. \hfill(8) \\
\\
\par As can be seen in (8), by minimizing its denominator, we can obtain the maximum throughput for different packet lengths. We define the parameter $OVRHD=1/\lambda +23.7\bar n+23.2$ and plot it in Fig. 1. It is obvious that minimizing $OVRHD$ maximizes the throughput. However, in addition to minimizing $OVRHD$, we consider other issues below.
\begin{figure}[!t]
\begin{center}
\epsfxsize=8cm \leavevmode\epsfbox{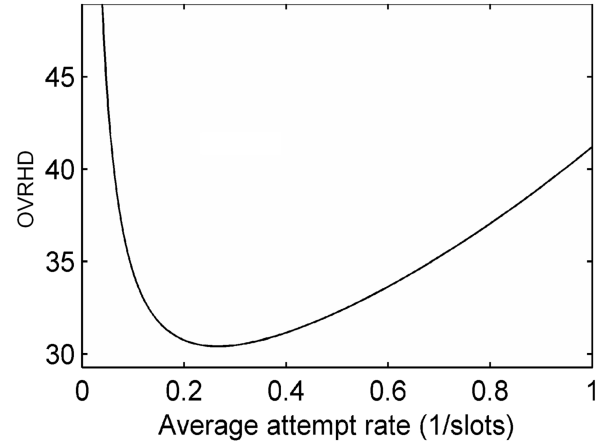} \caption{$OVRHD$ versus attempt rate.} \label{fig:1}
\end{center}
\end{figure}

\par According to (1), substitution of $M$ by $k_1M$ results in a value of $\bar b$ that corresponds to the case when we use $\lambda /k_1$ instead of $\lambda$. Consequently, given a fixed desired attempt rate, errors in the estimation of $M$, rapid fluctuations in its value, and variations in $\bar b$ lead to frame transmissions with a different attempt rate, i.e., the actual attempt rate of the network. For $k_1$ greater than 1, stations transmit with an attempt rate less than the desired value, and for a $k_1$ less than 1, stations transmit with an attempt rate greater than the desired value. It is observable in Fig. 1 that changing the attempt rate can increase $OVRHD$. Therefore, we should fix the attempt rate to a value that makes the system less sensitive to factors such as estimation errors. As can be seen in Fig. 1, variation of $OVRHD$ is greater when the attempt rate is less than 0.26 (1/slots). If we fix the attempt rate to a value less than 0.26 (1/slots), errors in the estimation of $M$ (specifically overestimation of $M$) may result in more variations in $OVRHD$ than when the attempt rate is greater than 0.26 (1/slots). Considering Fig. 1, the values between 0.26-0.8 (1/slots) are good choices for the attempt rate. In addition, as can be seen from Fig. 2, the mean access delay, obtained by $\bar d\!=\!\bar n(\bar {cw}+EIFS+tRTS)+\bar {cw}\!=\!\bar n(1/\lambda +26.2)+1/\lambda $ , rapidly increases for attempt rates smaller than 0.2 (1/slots). Mean access delay is the time interval between the start of backoff procedure for transmission of a frame and the beginning of the successful transmission of its first bit.
\par It can be inferred from $\bar d = \bar n(1/\lambda +26.2)+1/\lambda $ that it is possible to reduce the access delay of particular stations or the packets of specific stations by increasing their $\lambda$ in a constant $\bar n$. Consequently, different levels of QoS can be provided for nodes or applications by changing their mean access times. However, it should be noted that we need to decrease $\lambda$ sufficiently for some other stations or applications if we are interested in fixing the average attempt rate of the network and the average number of collisions between two consecutive successful transmissions. Here, we explain the mechanism for QoS differentiation by tuning the station attempt rates using an example. Suppose that we have a WLAN with $M/2$ stations transmitting their frames with an average backoff time of $0.25\bar{b}$ and $M/2$ stations transmitting with an average backoff time of $1.75\bar{b}$. The average backoff time of the network is $\bar b$ and the average attempt rate will be $\lambda =M/\bar{b}$. Additionally, the attempt rate of these two groups of nodes will be $2\lambda$ and $2\lambda/7$, respectively.
\begin{figure}[t]
\begin{center}
\epsfxsize=8cm \leavevmode\epsfbox{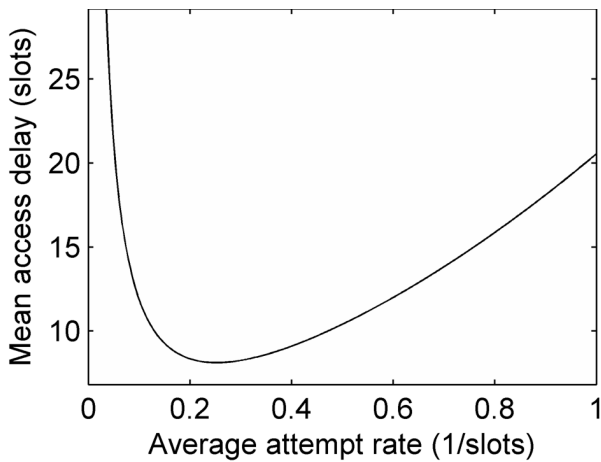} \caption{Mean access delay versus attempt rate for the case with the RTS/CTS mechanism.} \label{fig:2}
\end{center}
\end{figure}
\par Another factor that can be considered when selecting the attempt rate is stability. The Laplace transform of the mean access delay is expressed as\\
\\
$D^*(s)=N_c(CW^*(s)CF^*(s)e^{-(EIFS)s})CW^*(s)$, \hfill(9) \\
\\
where $CW^*(s)$ and $CF^*(s)$ denote, respectively, the Laplace transform of the probability density functions associated with $CW$ and $CF$. The random variable $CW_i$ denotes the number of idle slots before the $i$th collision or successful transmission, and $CF$ is the random variable indicating the size of a collided frame. The distribution of $CF$ is the same as that of $X'$ and is given. The average of $X'$ is $\bar x'$.
\par Using (9) with (2) and substituting $CW^*(s)$ by $exp(-s/\lambda )$ (where $1/\lambda $ is the average number of idle slots before each transmission attempt), we obtain\\
\\
$D^*(s)={\lambda e^{-\lambda }\over(1-e^{-\lambda })e^{s/\lambda }-(1-e^{-\lambda }-\lambda e^{-\lambda })e^{-26.2s}}$.\hfill(10) \\
\\
Furthermore, $D^*(s)$ has a real pole in the left region of the s-plane. The distance of this pole from the imaginary axis versus attempt rate is plotted in Fig. 3.
\begin{figure}[t]
\begin{center}
\epsfxsize=8cm \leavevmode\epsfbox{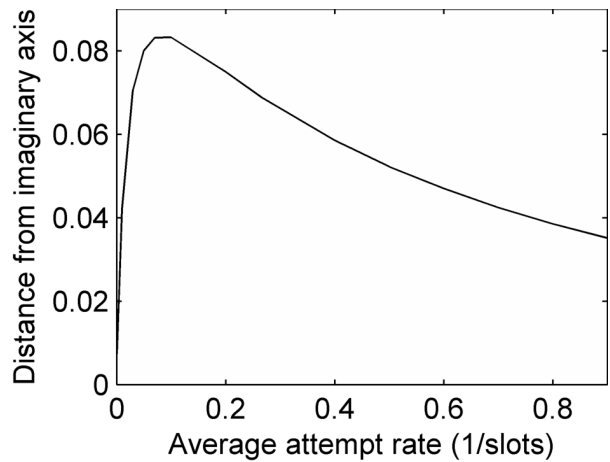} \caption{Distance of the pole versus attempt rate for the case with the RTS/CTS mechanism.} \label{fig:3}
\end{center}
\end{figure}
It can be seen in Fig. 3 that for very small and large attempt rates, the distance of the pole from the imaginary axis is very small. We should consider this fact in the selection of $\lambda $. The robustness of the protocol for five different values of $\lambda $ is presented in Table 2. The maximum allowable variation of the mean access delay is 10\% when calculating the maximum and minimum tolerable $\tilde {M}/M$. Choosing a $\lambda $ greater than 0.26 (1/slots), decreases the throughput while increasing the mean access delay. It also enhances the protocol robustness to the overestimation of $M$ while degrading the protocol robustness to the underestimation of $M$. The degradation in this case is rather low. Considering the above points, we adopt 0.7 (1/slots) as the attempt rate. However, as we have previously stated, values between 0.26-0.8 (1/slots) are appropriate choices for the average transmission attempt rate.
\begin{table}
\caption{Effects of different attempt rates (with the RTS/CTS mechanism)}
\label{tab:tab2}
\begin{center}
{\small \begin{tabular}{|c|c|c|c|}\hline
$\lambda $ & Mean & Maximum & Minimum \\
(1/slots) & access & tolerable & tolerable \\
 & delay & $\tilde M/M$ & $\tilde M/M$ \\
 & (slots) &  &  \\ \hline
0.1 & 12.06 & 1.25 & 0.83 \\
0.4 & 9.09 & 3 & 0.85 \\
0.5 & 10.39 & 4.5 & 0.89 \\
0.7 & 13.81 & 9.6 & 0.92 \\
1 & 20.08 & 13.8 & 0.97 \\ \hline
\end{tabular}}
\end{center}
\end{table}

\subsection{Determination of the attempt rate for the case without the RTS/CTS mechanism}
\par In this subsection, concerns similar to the case with the RTS/CTS mechanism are addressed when choosing the transmission attempt rate. When the RTS/CTS mechanism is not used, the equation for throughput is\\
\\
$T = {\bar x'\over\bar x'+8.6+1/\lambda +15.7\bar n+\bar n.\bar x'}$, \hfill(11) \\
\\
In (11), $OVRHD_2=8.6+1/\lambda +15.7\bar n+\bar n.\bar x'$ is an indicator of the overhead required for the transmission of a packet. Therefore, it is evident that we should have $\bar x'$ for minimizing the overhead. Since the goal of the backoff tuning algorithm is to guarantee a balance between the average collision cost $E[coll]$ and the average length of the idle period $E[idle]$ in a frame service time [9], for a frame service time we should have\\
\\
$E[coll]=E[idle]$. \hfill(12) \\
\par Because the average number of collisions in a frame service time is $\bar n$ and the average length of a collided frame is $\bar {cf}$, we have\\
\\
$E[coll]=\bar n.\bar {cf}$ and\\
\\
$E[idle]=\bar n.EIFS+(\bar n+1).\bar {cw}+DIFS+SIFS$.\\
\\
Hence, the value of $\bar x'$ that satisfies (12) is obtained as a function of $\lambda $. Thus\\
\\
$\bar x'=18.2+(1+1/\bar n)/\lambda +3/\bar n$. \hfill(13) \\
\par Now, we have $OVRHD_2$ as a function of $\lambda $. When $\lambda $ is equal to 0.31 (1/slots), $OVRHD_2$ is at a minimum and in this case, we have $\bar x'$ = 58 (slots) from (13).
\par The mean access delay when no utilizing RTS/CTS is\\
\\
$\bar d = \bar n(\bar {cw}+EIFS+\bar x')+\bar {cw} = \bar n(1/\lambda +18.2+\bar x')+1/\lambda $. \hfill(14) \\
\\
Substituting different attempt rates and their corresponding optimal $\bar x'$ in (14), we calculated the mean access delay. Results are shown in Fig. 4.
\begin{figure}[t]
\begin{center}
\epsfxsize=8cm \leavevmode\epsfbox{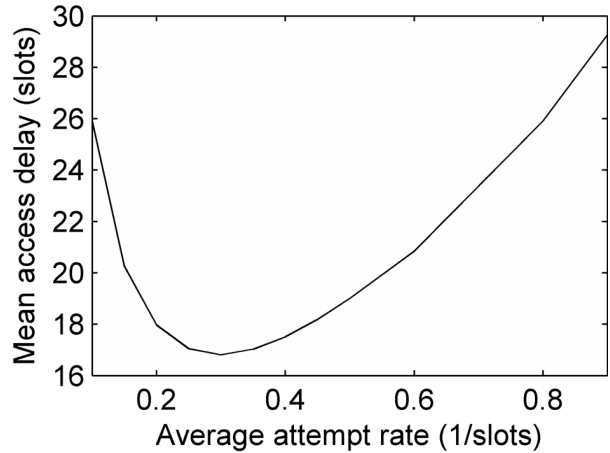} \caption{Mean access delay versus attempt rate.} \label{fig:4}
\end{center}
\end{figure}
We suggest that values between 0.3-0.7 (1/slots) are good attempt rate choices for acceptable performance in the presence of attempt rate variations. The mean access delay is minimal for an attempt rate of 0.31 (1/slots). However, protocol performance is more sensitive to attempt rate variations than situations with greater attempt rates.
In addition to the performance degradation caused by the variation in the attempt rate, the stability of the network should be considered when determining the attempt rate. For the case without the RTS/CTS mechanism, the Laplace transform of the mean access delay is\\
\\
$D^*(s)={\lambda e^{-\lambda }\over(1-e^{-\lambda })e^{s/\lambda }-(1-e^{-\lambda }-\lambda e^{-\lambda })e^{-(18.2+\bar x')s}}$.\hfill(15) \\
\par Similar to the RTS/CTS case, $D^*(s)$ has a real pole near the origin. The distance of this pole from the imaginary axis versus attempt rate is depicted in Fig. 5. In this figure, the average packet length satisfies (13) for each $\lambda $. From Fig. 5, we infer that the maximum stability of the system is at the attempt rate of 0.45 (1/slots). In addition, it can be seen in this figure that the distance of the pole from the imaginary axis for $\lambda $ = 0.7 (1/slots) is equal to that of $\lambda $ = 0.3 (1/slots). Consequently, we focus on values of $\lambda $ that are between 0.3-0.7 (1/slots) for choosing the transmission attempt rate. The distance of the pole from the imaginary axis versus $\lambda $ is plotted in Fig. 6 for various packet lengths. In Fig. 6, each packet length is calculated by (13) using four different attempt rates: 0.31, 0.45, 0.55, and 0.7 (1/slots). Packet lengths are in slots in this figure.
\begin{figure}[h!tb]
\begin{center}
\epsfxsize=8cm \leavevmode\epsfbox{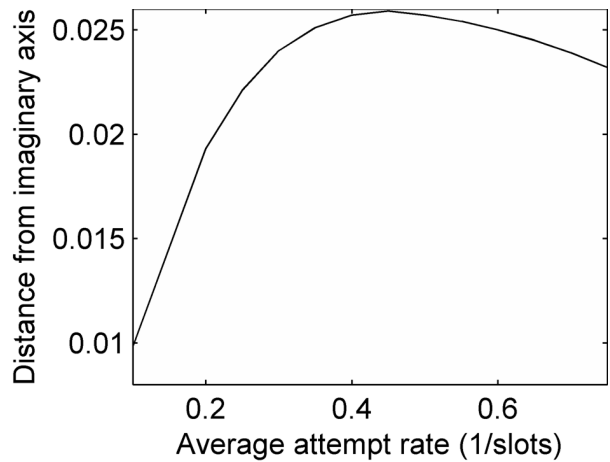} \caption{Distance of the pole from the imaginary axis versus attempt rate.} \label{fig:5}
\end{center}
\end{figure}
\begin{figure}[h!tb]
\begin{center}
\epsfxsize=8cm \leavevmode\epsfbox{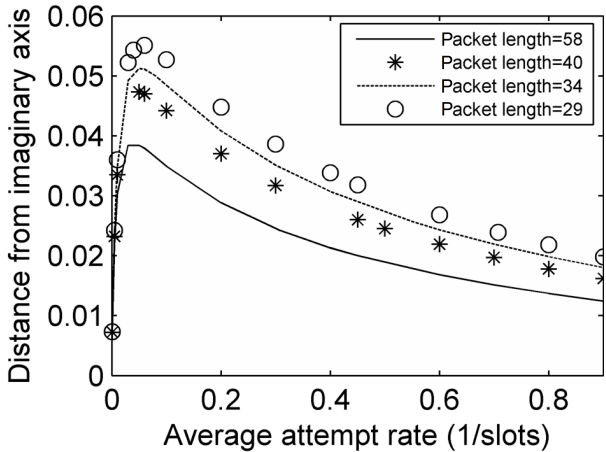} \caption{Distance of the pole from the imaginary axis versus attempt rate for different packet lengths (packet length in slots).} \label{fig:6}
\end{center}
\end{figure}
\par As can be seen in Fig. 6, reducing the packet length (the effect of increasing $\lambda $) enhances stability. However, the cost of this enhancement is throughput degradation and a larger mean access delay. In other words, smaller delay and larger throughput results in less robustness to variations in the attempt rate and lowers stability. The results of choosing five different values for $\lambda $ are presented in Table 3. Similar to the case with the RTS/CTS mechanism, the maximum allowable variation of the mean access delay is 10\% when calculating the maximum and minimum tolerable $\tilde {M}/M$. In the case without the RTS/CTS mechanism, $\lambda $ = 0.55 (1/slots) is selected as the attempt rate and therefore the optimal packet length that satisfies (13) is 34 (slots).
\begin{table*}
\caption{Effects of different attempt rates (without the RTS/CTS mechanism)}
\label{tab:tab3}
\begin{center}
{\small \begin{tabular}{|c|c|c|c|c|c|c|}\hline
 &  &  &  &   &  Maximum &  Minimum \\
$\lambda $ & $\bar x'$ & $\bar d$ & ${\bar d \over {\bar x'}} \times 100$ & $T$ &  tolerable &  tolerable \\
 (1/slots) & (slots) & (slots) & & & $\tilde M/M$ & $\tilde M/M$ \\ \hline
0.31 & 58 & 16.84 & 31.95 & 70.34 \% & 4.81 & 0.88 \\
0.45 & 40 & 18.11 & 45.28 & 61.10 \% & 7.90 & 0.90 \\
0.55 & 34 & 19.82 & 58.30 & 55.76 \% & 11.00 & 0.91 \\
0.6 & 32 & 20.87 & 65.19 & 53.41 \% & 12.50 & 0.92 \\
0.7 & 29 & 23.49 & 81.00 & 48.89 \% & 16.80 & 0.92 \\ \hline
\end{tabular}}
\end{center}
\end{table*}
\par Succinctly, when the RTS/CTS mechanism is not used, stations must fix their attempt rates to 0.55 (1/slots) and use an average packet length of 34 (slots) for their transmissions. The mean access delay and throughput for a packet length of 34 (slots) are plotted in Fig. 7.
\par Similar to the RTS/CTS case, we can decrease the access delay of arbitrary stations or applications in order to enhance their priority in accessing the medium (to provide a higher level of QoS) by increasing their transmission attempt rates. As can be observed from (14), increasing $\lambda$ reduces the mean access delay. In addition, when increasing the attempt rate for some stations or applications, it is necessary to adequately decrease the attempt rate of other stations or applications to maintain the fixed average attempt rate of the network.  
\begin{figure*}[!t]
\begin{center}
\epsfxsize=16cm \leavevmode\epsfbox{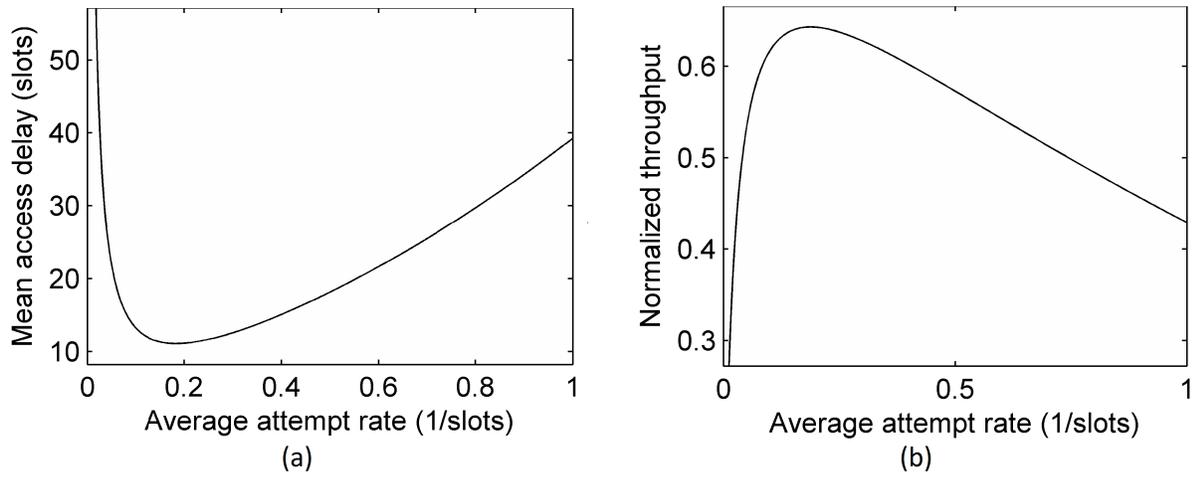} \caption{(a) Mean access delay versus attempt rate for a packet length of 34 (slots). (b) Normalized throughput versus attempt rate for a packet length of 34 (slots).} \label{fig:7}
\end{center}
\end{figure*}
\vspace{10pt}
\section{\uppercase{ANALYTICAL AND SIMULATION RESULTS}}
\label{sec:result}
\par In this section, we investigate the performance of the proposed ABTMAC protocol. We used the iterative method introduced in [1] to calculate the attempt rate in a network that uses the legacy DCF. Performance evaluation is performed analytically using the analytical model introduced and verified in [1] by substituting the corresponding parameters of legacy DCF and ABTMAC in the related equations. In addition, OPNET simulation results verify that our proposed MAC protocol successfully enhances the network performance.
\subsection{Analytical results}
\par In this subsection, packet lengths are given in slots. However, it may be helpful to know that the length of a 10 (slots) packet is 25 bytes given our current settings. The results of applying an attempt rate of 0.7 (1/slots) when using the RTS/CTS mechanism compared to legacy DCF are illustrated in Figs. 8a and 8b. For the case without the RTS/CTS mechanism, the normalized throughput and the mean access delay of systems using the legacy DCF and ABTMAC are depicted in Figs. 8c and 8d, respectively. As can be seen in these figures, performance improvement obtained by ABTMAC is more in larger values of $M$.
\begin{figure*}[!t]
\begin{center}
\epsfxsize=16cm \leavevmode\epsfbox{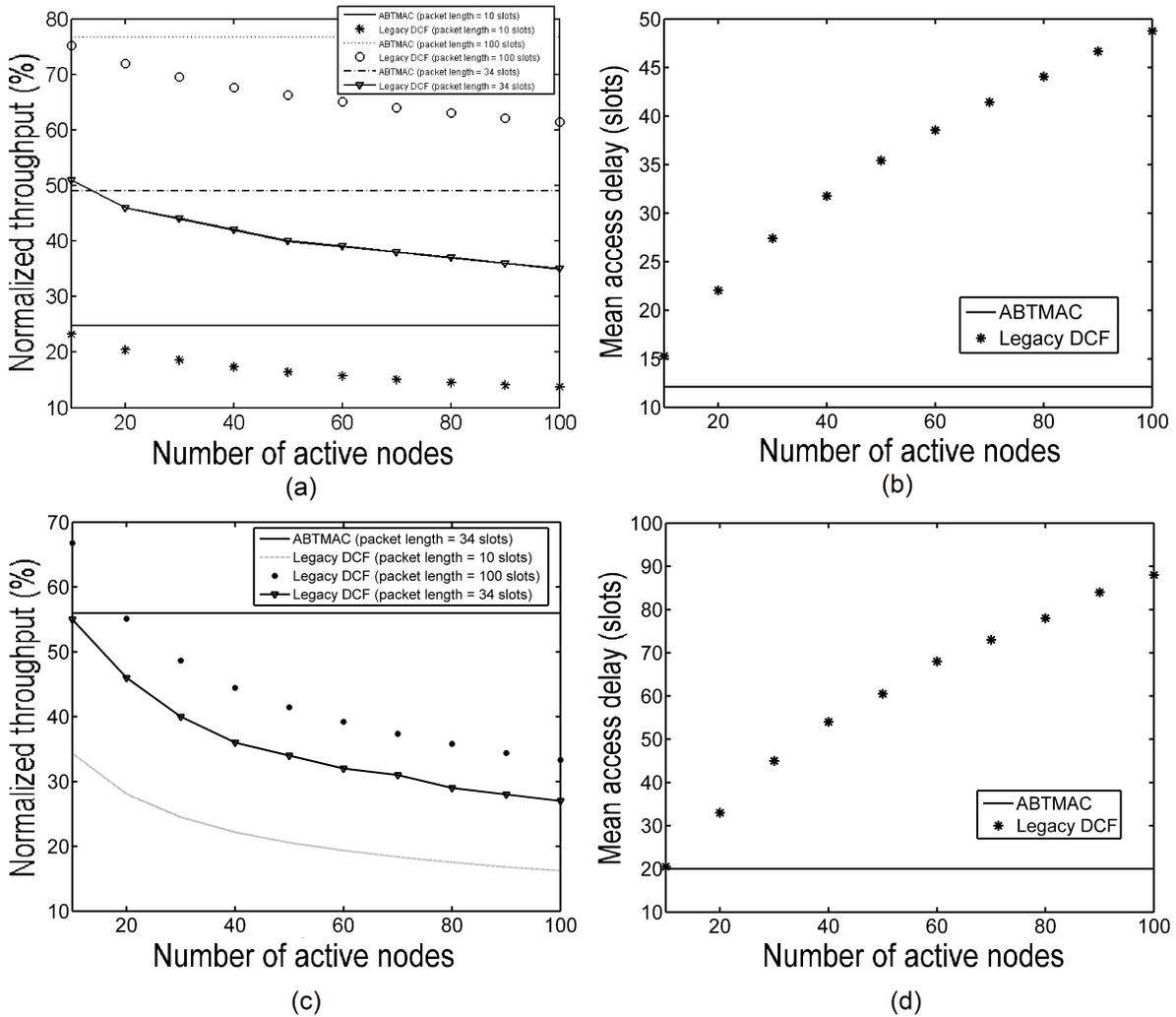} \caption{(a) Throughput versus $M$ (RTS/CTS mode). (b) Mean access delay versus $M$ (RTS/CTS mode). (c) Throughput versus $M$ (without RTS/CTS). (d) Mean access delay versus $M$ (without RTS/CTS, packet length = 34 slots).} \label{fig:8}
\end{center}
\end{figure*}
\begin{figure*}[t]
\begin{center}
\epsfxsize=16cm \leavevmode\epsfbox{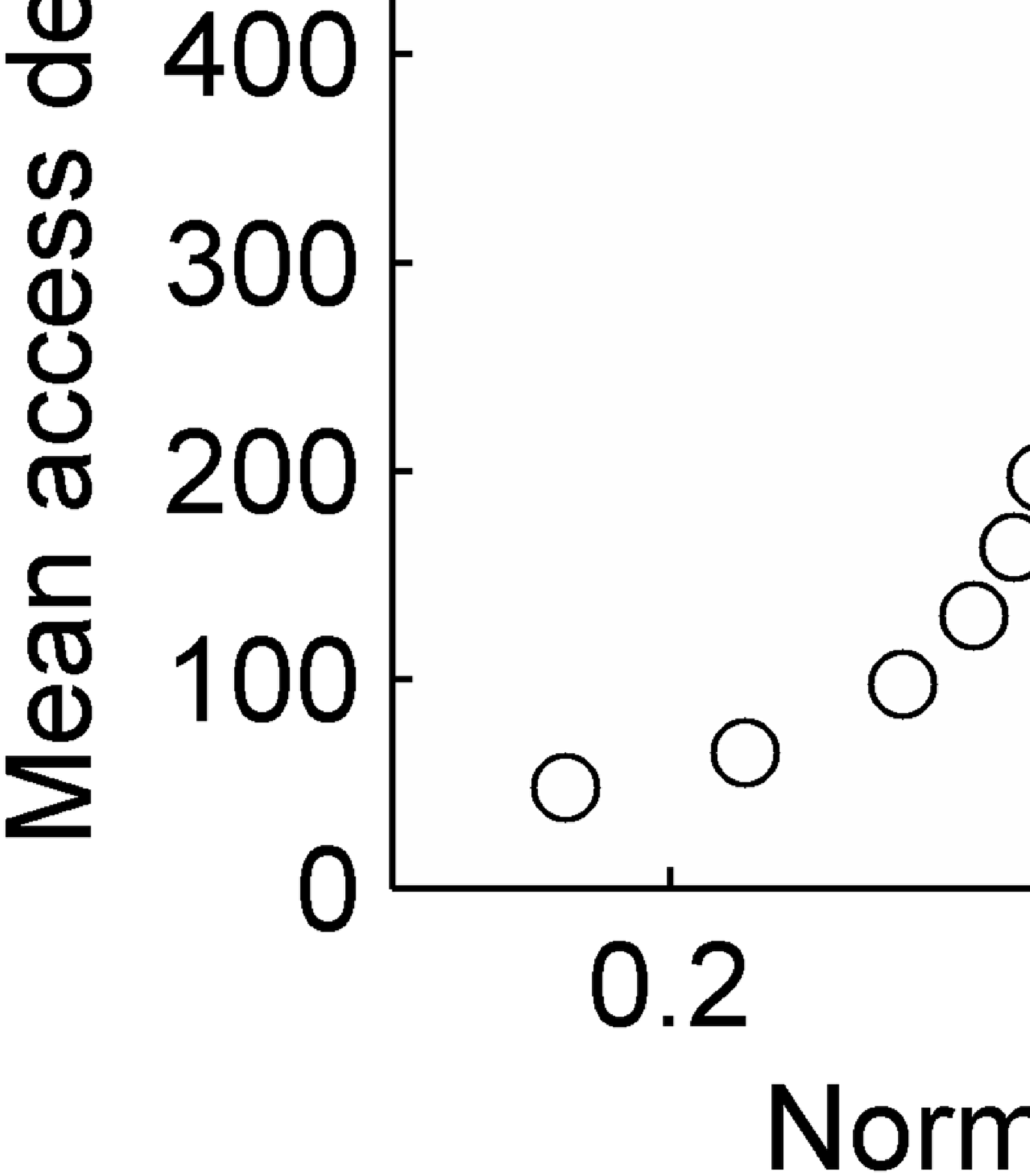} \caption{(a) Mean access delay for the IEEE 802.11 standard and our proposed MAC protocol in different network configurations (without RTS/CTS). (b) Distance of the pole from the imaginary axis versus packet length in various network configurations for the IEEE 802.11 standard and our proposed MAC protocol (without RTS/CTS).} \label{fig:9}
\end{center}
\end{figure*}
\par Fig. 9a shows the mean access delay versus the normalized throughput for ABTMAC and legacy DCF. In legacy DCF, achieving a larger throughput increases the mean access delay for larger values of $M$. However, applying our proposed MAC protocol leads to a larger throughput without greater mean access delay as the number of active nodes grows. The stability of the IEEE 802.11 standard and ABTMAC is shown in Fig. 9b. It should be noted that the performance of ABTMAC is independent of $M$.
\par The average number of collisions during one frame service time and collision probabilities are shown in Figs. 10 and 11, respectively, for the IEEE 802.11 standard and our proposed ABTMAC with and without the RTS/CTS mechanism. Although the analytical results presented in these figures show that the number of collisions between two consecutive successful transmissions and the collision probability remains constant when $M$ increases, stations are not able to estimate the number of active nodes accurately in a practical implementation. In addition, the value of the backoff time chosen from the CW is random and the average backoff time of the network is different from its theoretical value for a certain attempt rate. Therefore, practical values for $\bar n$ and collision probability will be greater than the analytical results presented in Figs. 10 and 11.
\begin{figure}[h!tb]
\begin{center}
\epsfxsize=8cm \leavevmode\epsfbox{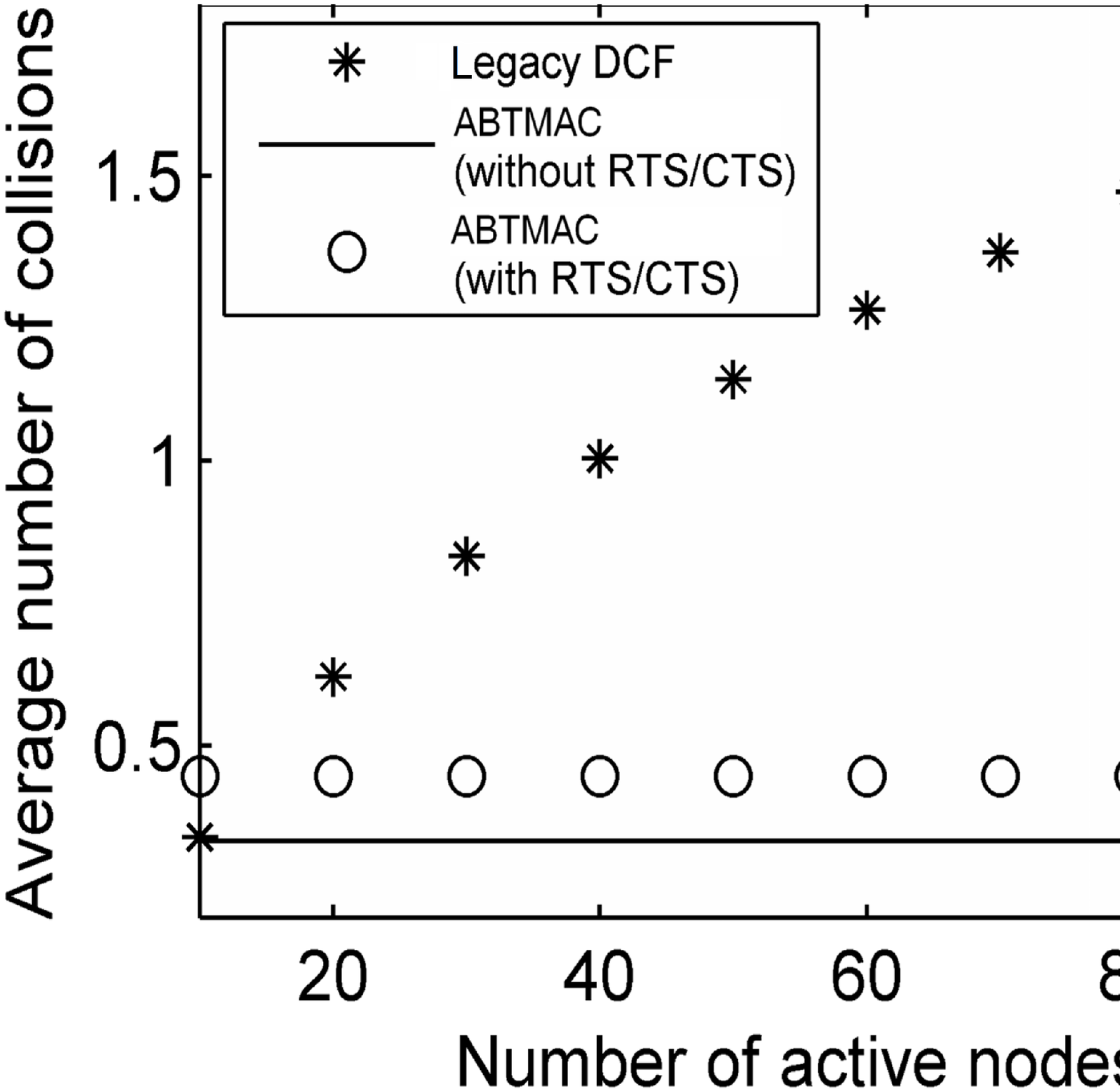} \caption{Average number of collisions versus $M$.} \label{fig:10}
\end{center}
\end{figure}
\begin{figure}[h!tb]
\begin{center}
\epsfxsize=8cm \leavevmode\epsfbox{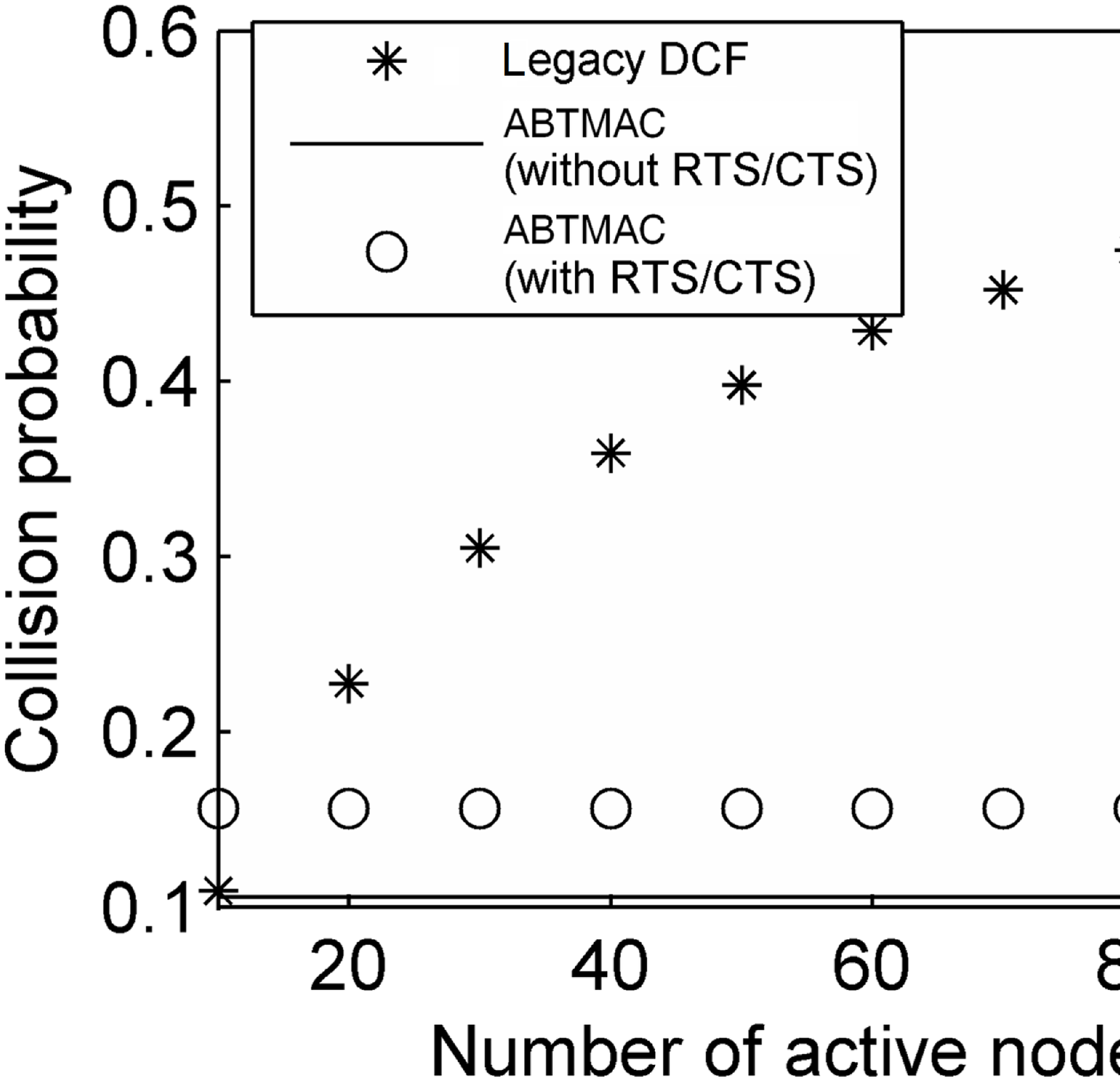} \caption{Collision probability versus $M$.} \label{fig:11}
\end{center}
\end{figure}
\par The utilization rate of slots (Slot Utilization) observed on the channel by each station is a simple and effective estimate of the channel congestion level. Slot utilization is defined as follows [4]\\
\\
$S\_U={number\ of\ busy\ slots \over number\ of\ available\ slots}$.\\
\\
The value of $S\_U$ is 0.60 when $M$ is 100 for legacy DCF with the RTS/CTS mechanism. By utilizing ABTMAC, it is 0.76 for all $M$ (the packet length is 34 (slots) in both cases). When the RTS/CTS mechanism is not used and $M$ is equal to 100, $S\_U$ is 0.73 for legacy DCF (with a packet length of 34 (slots)), and 0.8 for our proposed MAC protocol.
\subsection{Simulation study}
\par To evaluate the performance of ABTMAC, we ran simulations using OPNET. There is File Transfer Protocol (FTP) traffic over Transmission Control Protocol (TCP) in an ad hoc network and the number of nodes is 100. Among the variants of TCP, TCP Reno was used in this evaluation. Filesize was set at 1000 bytes. Profile configuration was as follows: The request start time for a file was uniformly distributed between 100-3400 sec. The number of repetitions was constant and equal to 3. In addition, the duration of requests was constant and equal to 10 sec. Both cases with and without the RTS/CTS mechanism were considered in our simulations. The data rate was 1 Mbps. Dynamic Source Routing (DSR) was used as routing protocol, and nodes were without mobility. Nodes were distributed over a 210 m $\times $ 210 m square area, and their transmission ranges were 300 m. $CW_{min}$ should be equal to 92 for an attempt rate of 0.55 (1/slots), and 72 for an attempt rate of 0.7 (1/slots) in a network with 100 active stations implementing ABTMAC. Simulation time was one hour (60 minutes). Finally, $CW_{min}$ was 15 in IEEE 802.11g. IEEE 802.11g is compared to ABTMAC, MFS, and AOB in the remainder of this section.
\par Results were measured using the facilities of the simulation software. WLAN throughput, defined in OPNET as the number of bits sent out from the MAC layer over the total number of bits sent to the MAC layer, was selected as our performance measure. The average of throughput was calculated for all the nodes in the network. The average throughput is shown in Figs. 12 and 13. These figures show the average of throughput up to an arbitrary simulation time for all the nodes in the network. Simulation time in minutes is specified by the x-axis in Figs. 12 and 13. The fragmentation threshold was 85 bytes (34 slots), and $CW_{min}$ was 92 for the case without RTS/CTS and 72 for the RTS/CTS case. There was no limitation on the packet length for the case with RTS/CTS, and we set the fragmentation threshold to 1024 bytes. Fig. 12 shows the throughput of the legacy DCF and ABTMAC, while Fig. 13 illustrates the throughput of the MFS and AOB schemes.
\begin{figure}[!t]
\begin{center}
\epsfxsize=8cm \leavevmode\epsfbox{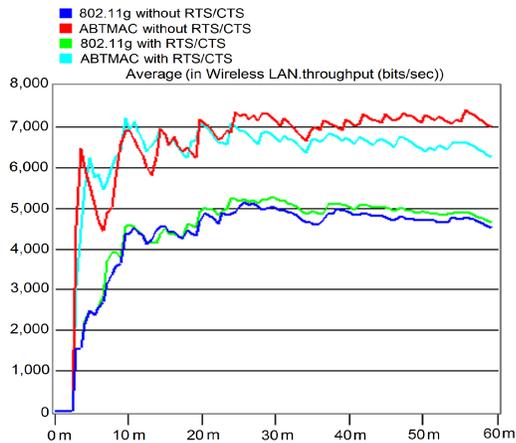} \caption{Average throughput for IEEE 802.11g and ABTMAC.} \label{fig:12}
\end{center}
\end{figure}
\begin{figure}[!t]
\begin{center}
\epsfxsize=8cm \leavevmode\epsfbox{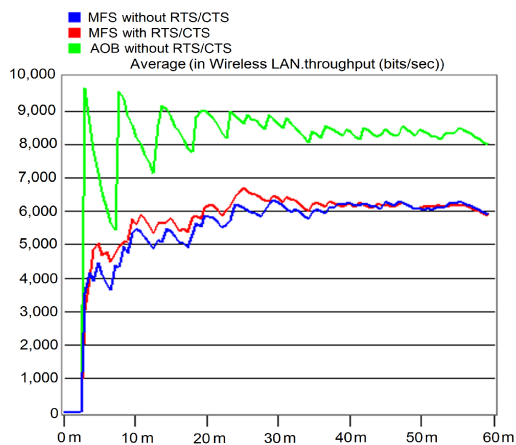} \caption{Average throughput for MFS and AOB.} \label{fig:13}
\end{center}
\end{figure}
\par WLAN throughput is enhanced by using the attempt rates specified in Section III. It can be seen in Fig. 12 that we have 38.9\% and 53.8\% throughput enhancement with and without RTS/CTS, respectively, when using ABTMAC instead of the legacy DCF. It is observable from Figs. 12 and 13 that ABTMAC outperforms MFS in the case without the RTS/CTS, and their performance is almost identical to the case with RTS/CTS. As previously stated, AOB does not consider the RTS/CTS mechanism. Using the method described in AOB mechanism [4], we obtain that stations should transmit their frames with a probability of 0.001798 in each slot time when the packet length is 34 slots and there are 100 active stations in the network. This probability of transmission leads to an attempt rate of 0.1801 (1/slots) that conforms to the results presented in Fig. 7. As pointed out in Section III, our proposed MAC protocol works at an attempt rate near the point that maximizes the throughput (or minimizes the access delay). Therefore, simulation results show that the performance enhancement of AOB is higher than ABTMAC in the case without the RTS/CTS mechanism. However, based on the analytical results presented in Fig. 7, the performance of AOB is sensitive and may degrade when system parameters change. The delay variation in the network, obtained from Fig. 7, when the number of active nodes rapidly becomes 3 times greater is 42.1\% and 81.8\% for ABTMAC and AOB, respectively. Since it is not possible to analyze MFS by the analytic model used in this paper, ABTMAC is compared to AOB in terms of the delay variation. The delay variation in ABTMAC is significantly smaller than AOB, and this is an advantage of the method.
\par Finally, we evaluated the performance of the network when the estimation of the number of active nodes is inaccurate and also when stations use packet lengths different from the optimal value obtained in Section III. Results are shown in Tables 4 and 5. We infer from Table 4 that the impact of underestimating $M$ is higher than its overestimation. Underestimating $M$ reduces the throughput by 11.9\% and 22.2\% in the cases with and without RTS/CTS, respectively. Overestimating $M$ in the case without the RTS/CTS mechanism decreases the throughput by 4.4\%. In the RTS/CTS case, overestimating $M$ leads to an actual attempt rate that reduces the $OVRHD$ and therefore we have 3\% throughput improvement. Based on the simulation results presented in Table 5, using a packet length of 2048 bytes, which is greater than the optimal value, decreases the throughput by 8\%.
\begin{table}
\caption{Effect of number of active nodes estimation errors on performance}
\label{tab:tab4}
\begin{center}
{\small \begin{tabular}{|c|c|c|}\hline
 Estimated & Throughput & Throughput \\
$M$ & (bps) & (bps) \\
  & (without RTS/CTS) & (with RTS/CTS) \\  
\hline
50 & 5440 & 5500\\
100 (without error) & 7000 & 6250 \\
150 & 6690 & 6438 \\ \hline
\end{tabular}}
\end{center}
\end{table}
\begin{table}
\caption{WLAN throughput for different packet lengths}
\label{tab:tab5}
\begin{center}
{\small \begin{tabular}{|c|c|}\hline
 Packet & Throughput \\
length & (bps) \\
 (Byte) &  \\  
\hline
85 (optimal value) & 7000\\
256 & 6650 \\
1024 & 6490 \\
2048 & 6440 \\ \hline
\end{tabular}}
\end{center}
\end{table}
\vspace{10pt}
\section{\uppercase{CONCLUSION}}
\label{sec:conc}
\par In this paper, we introduced ABTMAC, a MAC protocol based on IEEE 802.11 DCF. In ABTMAC, a higher protocol capacity in comparison with legacy DCF is achieved by tuning the backoff window size and using an optimal packet length (when the RTS/CTS mechanism is not used). Tuning the backoff window size is done via fixing the transmission attempt rate of stations. We determined appropriate values for the attempt rate in the cases with and without using the RTS/CTS mechanism. Our proposed ABTMAC protocol is backward compatible with legacy DCF. Different levels of QoS can be guaranteed for different users by specifying appropriate attempt rates. In addition to the analytical results, the performance of the proposed MAC protocol was evaluated through simulations. The results showed that ABTMAC can significantly outperform IEEE 802.11 DCF and also offers satisfactory performance in comparison to other similar approaches.
\bibliographystyle{jcn}


\epsfysize=3.2cm
\begin{biography}{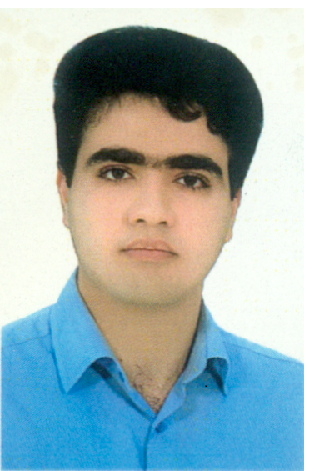}{Amin Jamali} received a B.S. degree in electrical engineering from Isfahan University of Technology, in 2004, and an M.S. degree in electrical engineering from Amirkabir University of Technology, in 2007. Currently, he is working towards the Ph.D. degree in electrical engineering at Isfahan University of Technology. His research interests include wireless networks, network security, network survivability, and intrusion detection.
\end{biography}

\epsfysize=3.2cm
\begin{biography}{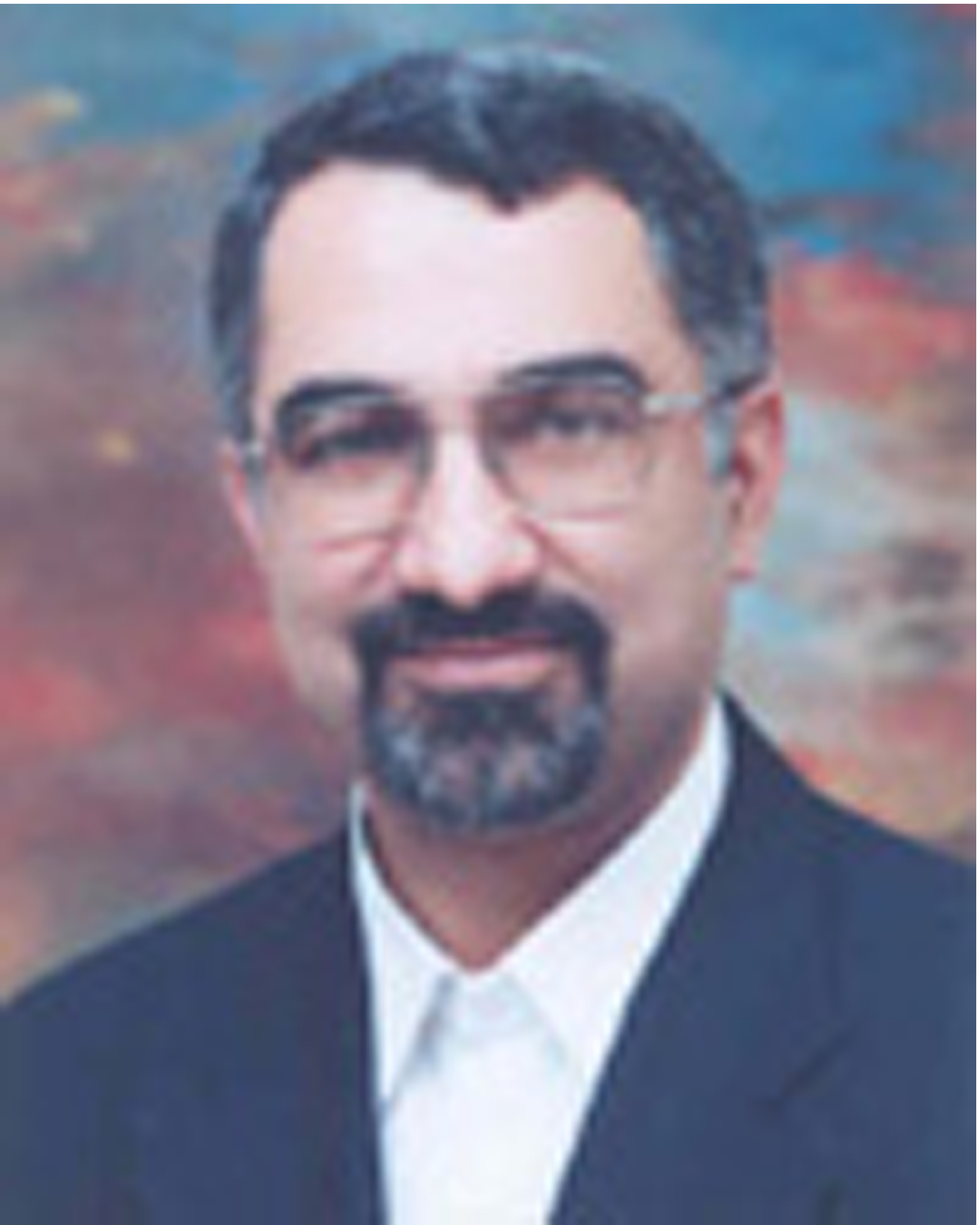}{Seyed Mostafa Safavi Hemami} received his B.S. and M.S. degrees in electrical engineering in 1986 and 1987, respectively from Technological University in Cookeville, Tennessee, USA. He obtained his Ph.D. degree in Information Technology (Communication) from George Mason University in Virginia, USA in 1993. He is currently an Associate Professor at Amirkabir University of Technology in Tehran, Iran, where he teaches High Speed Networks, Satellite Communication, and Queuing Theory for Communication Systems among other courses.
\end{biography}

\epsfysize=3.2cm
\begin{biography}{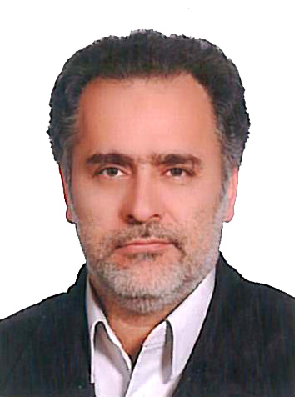}{Mehdi Berenjkoub} received a Ph.D. degree from the Department of Electrical and Computer Engineering, Isfahan University of Technology in 2000. The subject of his dissertation was two-party key distribution protocols in cryptography. He started his work in the same department as an assistant professor from that time. Graduate courses presented by him include Fundamentals of Cryptography, Cryptographic Protocols, Network Security, and Intrusion Detection. He has supervised more than a dozen M.Sc. students and Ph.D. candidates in related areas. He also was one of the founding members of the Iranian Society of Cryptology in 2001. He has continued his cooperation with the society as an active member. He along with his colleagues recently established a research group on Security in Networks and Systems in Isfahan University of Technology. He also is responsible for a newly established academic CSIRT in IUT. His current interested research topics are wireless network security and authentication protocols.
\end{biography}

\epsfysize=3.2cm
\begin{biography}{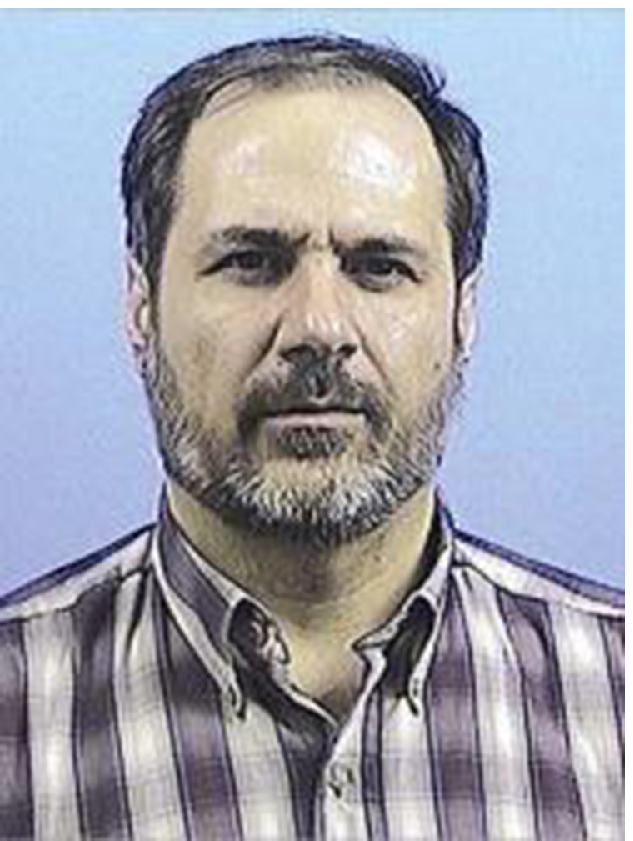}{Hossein Saidi} received B.S. and M.S. degrees in electrical engineering (Electronics and communication Eng.) from Isfahan University of Technology (IUT), in 1986 and 1989, respectively, and a Ph.D. degree in electrical engineering from Washington University in St. Louis, MO in 1994. From 1994 to 1995, he was a research associate at Washington University St. Louis, and in 1995 he joined the Electrical and Computer Engineering Department of IUT, where he is an Associate Professor. His research interests include high-speed networking, wireless networks, QoS guarantees, routing algorithms, and information theory.
\end{biography}

\end{document}